\input epsf.tex
\documentstyle[aasms]{article}
\def\etal{{\it et al. }}
\def\spose#1{\hbox to 0pt{#1\hss}}
\def\lsim{\mathrel{\spose{\lower 3.0pt\hbox{$\mathchar"218$}}
     \raise 2.0pt\hbox{$\mathchar"13C$}}}
\def\gsim{\mathrel{\spose{\lower 3.0pt\hbox{$\mathchar"218$}}
     \raise 2.0pt\hbox{$\mathchar"13E$}}}
\def\msun{{\rm\,M_\odot}}
\def\rsun{{\rm\,R_\odot}}

\begin{document}
\title{CLOSE ENCOUNTERS BETWEEN A NEUTRON STAR AND A MAIN-SEQUENCE STAR}
\author{Hyung Mok Lee}
\affil{Department of Earth Sciences, Pusan University, Pusan 609-735, Korea}
\affil{e-mail:hmlee@astrophys.es.pusan.ac.kr}
\author{Sungsoo S. Kim}
\affil{Institute for Basic Sciences, Pusan University, Pusan 609-735, Korea;
and \break Department of Physics \& Astronomy, University of California,
Los Angeles, CA 90024, U. S. A.}
\affil{e-mail:sskim@astrophys.es.pusan.ac.kr}
\author{and}
\author{Hyesung Kang}
\affil{Department of Earth Sciences, Pusan University, Pusan 609-735, Korea}
\affil{e-mail: kang@astrophys.es.pusan.ac.kr}

\begin{abstract}
We have examined consequences of
strong tidal encounters between a neutron star and a normal star using
SPH as a possible formation mechanism of isolated recycled pulsars in
globular clusters.
We have made a number of SPH simulations for close encounters between a
main-sequence star of mass ranging from 0.2 to 0.7 $\msun$ represented by
an n=3/2 polytrope and a neutron star represented by a point mass.
The outcomes of the first encounters are found to be dependent only on the
dimensionless parameter $\eta^{\prime} \equiv
(m/(m+M))^{1/2} (r_{\min}/R_{MS})^{3/2}(m/M)^{(1/6)}$, where m and M are
the mass of the main-sequence star and the neutron star, respectively,
$r_{\min}$ the minimum separation between two stars, and $R_{MS}$ the size
of the main-sequence star.
The material from the (at least partially) disrupted star forms a disk
around the neutron star.  If all material in the disk is to be acctreted onto
the neutron star's surface, the mass of the disk is enough to spin up the
neutron star to spin period of 1 ms.
\end{abstract}


\section{INTRODUCTION}
The cores of globular clusters have high stellar densities. 
Recent studies showed that the physical interactions
between stars are main driving force in determining the dynamical evolution
after the core collapse (e.g., Goodman 1988). The physical interactions include 
direct collisions between normal stars, tidal captures between a normal star
and a compact star (neutron star or white dwarf), and formation of binaries
via three-body processes.

Such interactions can lead to the formation of objects that are not common
in low stellar environments. Indeed globular clusters reveal high abundance 
of X-ray binaries and short period
pulsars. These systems are thought to be related to the physical interactions
between stars. X-ray binaries can be formed either by tidal capture between
a neutron star and a main-sequence star, or by the evolution of primordial
binary.

Another observational evidence for the high abundance of the physical
interactions in clusters is the enhancement of short-period pulsars.
A large number of short period pulsars have been detected in globular 
clusters. Unlike many pulsars in the solar neighborhood, cluster pulsars are
known to be predominantly short period ones.
Neutron stars are believed to be born with short spin periods ($\lsim$
0.1 sec) and strong magnetic fields ($10^{12} - 10^{13}$ G).
As a neutron star
evolves, its rotational energy is released through magnetic dipole radiation
which causes deceleration of the spin. Since the spin-down time scales for
pulsars with strong magnetic fields are of order $10^8$ years while the
production of neutron stars in globular cluster must have ceased long time
ago, present day population of short period pulsars in clusters must be
`recycled' pulsars, which has been spun up by accretion of mass.

Aside from the fact that the abundance of recycled pulsars in globular clusters
relative to ordinary stars is far greater than that for solar neighborhood, the 
cluster pulsars are preferably isolated (as opposed to in a binary system).
According to compilation of 558 pulsars by Taylor, Manchester, \& Lyne (1993),
19 out of 22 in clusters are {\it isolated} pulsars while the ratio is only 2
out of 11 in field pulsars. 
Therefore any theory for the recycled pulsars in globular clusters should 
explain both the high abundance and high ratio of isolated to binary pulsars.

Proposed mechanisms for producing isolated recycled pulsars in
dense stellar regions include close encounters between a neutron star and
a main-sequence star leading to a complete or partial disruption of the
main-sequence star (Krolik 1984, Lee 1992), and the three-body interactions
which can liberate a neutron star in the binary through various ways (see
Rappaport,Putney, \& Verbunt 1989 for summary).  In the present paper, as a
start of the study on the first mechanism, we examine the consequences of close
encounters between a neutron start and a main-sequence star.

During a close encounter involving a neutron star, the orbital energy of an
incident normal star can be dissipated and stored in its
interior by tidal force exerted by a neutron star.  Depending on the amount of 
energy deposited to the stellar envelope, one of the following outcomes are 
possible:
tidal capture of a normal star, or total or partial disruption of the normal
star. Based on SPH simulations, Davies, Benz, \& Hills (1992) find
that encounters involving a main-sequence star of $0.8 \msun$ can leave a
single object for $r_{\min} \lsim 1.75 \, R_{MS}$ and detached binaries
for $1.7 \, R_{MS} \lsim r_{\min} \lsim 3.5 \, R_{MS}$,
while encounters involving a red giant star leave a compact neutron
star-white dwarf binary for $r_{\min} \lsim
1.8 \, R_{RG}$ and a detatched binary for $1.8 \, R_{RG}
\lsim r_{\min} \lsim 2.5 \, R_{RG}$, where, $r_{\min}$ is the separation at
periastron passage.
In the present study, while Davies \etal concentrated on only one main-sequence
star mass, we perform SPH simulations of encounters between a $1.4 \msun$
neutron star and a main sequence star with a mass in the range 
between $0.2$ and 0.7 $\msun$ for various $r_{\min}$.
This extension will bring a full comparison between simulation
and theory, and will eventually provide more realistic information
to the study of dynamical evolution of globular clusters.

This paper is organized as follows.  In \S 2, we describe our simulations
of stellar encounters, and we discuss the consequences of the encounters
based on the simulations in \S 3.  Final section summarizes our major findings.

\section{SIMULATION}

We adopt an SPH code developed by Monaghan \& Lee (1994) for the encounters 
between a massive black hole
and a normal star which is modelled as an $n$=3/2 polytropic sphere.
We use 9185 SPH particles to model a main-sequence star and treats the 
neutron star as a 1.4 $\msun$ point mass that interacts with SPH particles 
only via gravitational forces.  We have chosen main-sequence star mass
$m$ to be 0.2, 0.3, 0.5, and 0.7 $\msun$ in
our simulations, and the main-sequence stars are assumed to have
a linear mass-radius relation (i.e. $R_{MS}/\rsun = m/\msun$).
For $R_{MS}$ of order $\rsun$ and relative velocity at
infinity, $v_\infty$, of order 10 ${\rm km \, s^{-1}}$, the relative
velocity $v_{min}$ at $r_{\min}$ becomes
\begin{eqnarray}
v_{min}^2 & = & {{\rm G}(M+m) \over r_{min}} + v_{\infty}^2 \\
&            \simeq & {{\rm G}(M+m) \over r_{min}},
\label{vmin}
\end{eqnarray}
where $M$ is the mass of the neutron star.  This means that $v_\infty$
is not an important parameter in determining the orbit of the star around
the neutron star, and in our simulations, $v_\infty$ is alwasys the
representative value in globular clusters, 10 ${\rm km \, s^{-1}}$.  Therefore
we assume that the relative orbit is parabolic for simplicity.
We made total of 16 simulations whose input parameters ($m$ and $r_{min}$)
are listed in Table 1.

For the purpose of illustration of the outcome of tidal encounters,
we have shown four snapshots of simulation m7p08 in Figure 1.  Small
dots are the SPH particles, and the curved line is the trajectory that
a main-sequence star would follow if it were a point mass.  In this figure, the
coordinate center is the location of the neutron star.
At $t$ = 1.8 h (relative to the start of the simulation), the shape of
the main-sequence star is already distorted like a Roche lobe which is
pointing off the neutron star slightly.  At $t$ = 2.2 h, the SPH particles
are spread out in two opposite directions so that the center of mass still
reside near the parabolic orbit, and some of the particles from the
main-sequence star start to spiral into the neutron star.  At $t$ = 3.3 h,
a disk is formed around the neutron star and is still connected
to the main body of the main-sequence star which has a long tail of
particles with relatively smaller velocities.  At $t$ = 4.2 h, the distortion
of main-sequence star has ceased and the main-sequence star is now
composed of a sphere with high density and two spiral arms.  The disk around
the neutron star is expanding in the orbital plane because the 
particles in the disk has large eccentricities which make them spend
much longer time at farther distances than the beginning of the
accretion ($t \simeq$ 3.3 h) when most particles are near their perigees.

\section{ANALYSIS}

The main purpose of the simulations in this study is to observe how much
mass of the main-sequence star can be transfered to the vicinity of
(or bound to) the neutron star in the process of close encounter,
as a function of $m$ and $r_{\min}$ (note that this mass may differ
from the actual mass that accretes onto the neutron star surface and
consequently spins up the star).  Also, we will attempt to predict the
future of the main-sequence star remnant, which can be parameterized by
its size, $r_{\min}$ at the second encounter,
and the period of its orbit.  The disk formed around the neutron star will
be discussed in connection to the spin-up of the neutron star.

\subsection{Mass Fractions}

To study separately the futures of the disk around the neutron star and of
the main-sequence star remnant, it is necessary to judge 
whether each particle will be bound to the neutron star, remain as a part of
the main-sequence star, or escape from the system.  Here
we employ the following criteria to distinguish among the above three
possibilities: \break
\leftline{\indent (A) if ($E_{NS} \leq 0$) and ($E_{NS}-E_{MS}
\leq 0$), the particle is bound to the neutron star; \break}
\leftline{\indent (B) elseif ($E_{MS} \leq 0$), bound to the main-sequence
star; \break}
\leftline{\indent (C) else, lost from the system. \break}
The relative energies $E_{NS}$ and $E_{MS}$ are defined as
\begin{eqnarray}
E_{NS} & = & {1 \over 2} {\bf v}^2 - {{\rm G} M \over |{\bf r}|} \\
E_{MS} & = & {1 \over 2} ({\bf v} - \langle {\bf v} \rangle )^2 - {{\rm G} m \over |{\bf r} - \langle {\bf r} \rangle |},
\end{eqnarray}
where ${\bf v}$ and ${\bf r}$ are the velocity and the distance
of a particle relative
to the neutron star and the brackets denote the averages taken over all
SPH particles.  Figure 2 is a reproduction of 
the last snapshot in Figure 1 with three different symbols.  Small dots,
large dots, and $\times$'s are the particles that satisfy the conditions A, B,
and C, respectively.  Note that there are two clear 
separations between small dots and large dots and between large dots and
$\times$'s.  The amounts of mass belonging to each criterion as
a function of time for simulation m7p08 are shown in Figure 3. These masses
are well defined because each line approaches to an asymptotic value
rather quickly.

However, the mass fractions by the above criteria for simulations m2p02 and
m7p05 do not approach to asymptotic values, because their
$r_{\min}$'s are so small and the main-sequence star experiences
so strong tidal force that the main-sequence star sprials into the
neutron star (or, more realistically, the latter spirals into the former)
in a relatively short
time instead of spending some time on the quite eccentric orbit as an
independent object.  Thus for these two simulations, it is impossible
to distinguish the first encounter and the subsequent one.  We do
not include these two simulations in the following mass fraction
anaylsis which is limited to the mass fractions of the first
encounters only.

The amounts of mass that satisfies criteria A, B, and C are plotted as a
function of $r_{\min}/R_{MS}$ in Figure 4 for all simulations except
simulations m2p02 and m7p06.
Each symbol denotes different main-sequence star mass (see the caption).
It is clear that at the same $r_{\min}/R_{MS}$, heavier main-sequence star
loses less fraction of its mass.  From the similarity of the curves for each
membership, one may expect a parameter that soley determines the mass fractions.
For a theoretical approach, we have first tried $\eta$ of Press \& Teukolsky
(1977).  The quantity $\eta$ measures the duration of periastron passage,
relative to the hydrodynamic time of the star, and is defined as
\begin{equation}
\eta \equiv {\left ( {m \over m+M} \right ) }^{1/2} {\left ( {r_{\min}
\over R_{MS} } \right ) }^{3/2}.
\end{equation}
However, it is found that $\eta$ does not describe the mass fractions in a
unified way.  Rather, we find that a small alternation to $\eta$ can make the
mass fraction be a function of only one parameter:  As shown in Figure 5,
the mass fractions of all simulations allign on a single curve if the
abscissa of Figure 4 is replaced with a new quantity
\begin{equation}
\eta^{\prime} \equiv {\left ( {m \over m+M} \right ) }^{1/2} {\left ( {r_{\min}
\over R_{MS} } \right ) }^{3/2} {\left ( {m \over M} \right )}^{1/6}.
\end{equation}
Theoretical base of the above alternation is as follows:
According to Press \& Teukolsky (1977), when the relative orbit
is assumed to be a parabola, the deposition of
orbital energy to stellar oscillations of the star with mass $M_1$ and the
radius $R_1$ due to the perturbing star with $M_2$ and $R_2$ can be
expressed by
\begin{equation}
\Delta E = \left ( {{\rm G}M_1^2 \over R_1} \right ){\left ( {M_2 \over M_1}
\right )}^2 \sum_{l=2,3,...} {\left ( {R_1 \over r_{\min}} \right )}^
{2l+2}T_l(\eta),
\end{equation}
where $l$ is the spherical harmonic index.
The dimensionless function $T_l(\eta)$ depends on the structure
of the star, and the values of $T_l(\eta)$ for the star with a polytropic
index $n=2/3$ has been calculated by Lee \& Ostriker (1986).  Then one could
expect that the amount of the mass detatched from a star
during an encounter, $\Delta m$, should be related to this $\Delta E$.
Furthermore, since the detatched mass is a function of $\eta^{\prime}$,
$\Delta E$ is expected to be a function of $\eta^{\prime}$ somehow as well.
Indeed we find that $\Delta E/m$ can be well expressed as a function of
$\eta^{\prime}$.  In Figure 6, $\Delta E/m$'s for three different main-sequence
star masses are plotted over $\eta$ and $\eta^{\prime}$.  This comparison
clearly shows that $\Delta E/m$ can be expressed more uniformly in terms of
$\eta^{\prime}$ than just $\eta$.

It would be worth while to relate $\Delta m/m$ with $\Delta E/m$.
Such relationship can be found in Figure 7.  Note that $\Delta m/m$ increases
with $\Delta E/m$ linearly until $\Delta m/m \sim 0.1$.  After that,
it increases quatratically, but the overall tendancy can be approximated
to be still linear.

The theoretical quantity $\Delta E$ can be obtained in our simulations by
finding the difference of the orbital total energies (kinetic + potential)
at the start and at the end of the simulation.  However, since this
calculation is meaningful only when there is almost no mass loss from the
main-sequence star, we compare $\Delta E$'s with
large $\eta^{\prime}$ only.  In Figure 8, one can
easily see that there is a good agreement between our simulations denoted
by several kinds of symbols and the theory denoted by the curves.

\subsection{The Main-Sequence Star Remnant}
All main-sequence stars in our simulations are captured by the neutron
stars and thus will experience subsequent encounters.  However, no
numerical method has been known to be able to handle such long period
of hydrodynamic simulations.  For this reason, we have performed the
simulations of
first encounters only and try to predict the perigee distance at
second encounter, $r_{\min ,2}$, and the period between first and
second perigees, $P$, from the last dumps of some of our simulations.
 
To estimate $r_{\min ,2}$ and $P$, we take $\bf{r}$ and $\bf{v}$
averages for the particles that satisfy the criterion B in the
previous subsection.  Since the quantities $r_{\min ,2}$ and $P$
approach to asymptotic values less quickly than the mass fractions,
we have extended simulations m7p08, m7p11, and m7p15 until they show
asymptotic values, which are listed in Table 2.
 
Approximately, simulation m7p15 has twice greater $r_{\min}$ than that
of simulation m7p08, but its $P$ is a thousand times longer implying
that $P$ has a strong dependency on $r_{\min}$.
 
In the above three simulations, $r_{\min ,2}$ is about 50 \% bigger
than $r_{\min}$.  From this fact, the amount of mass stripped during the
second encounter, $\Delta m_2$,
might be exptected to be not as much as that for the first encounter,
$\Delta m$, but one
should remind that the parameter that determines $\Delta m$, $\eta^
{\prime}$, is a function of $R_{MS}$ too.  Since the main-sequence
stars are expanded during encounters (see Fig. 9), $R_{MS}$ will have been
enlarged
at the time of second encounter resulting smaller $\eta^{\prime}$
and greater $\Delta m$ than would be with a constant $R_{MS}$.
Furthermore, as in case of simulation m7p15, even considerable changes
in the structure accompany.
Also, the rotational angular momentum of the main-sequence star transfered from
orbital angular momentum, which is another result of the tidal interaction, may
play an important role in determining the amount of mass loss during subsequent
encounters.  Thus we conclude that the prediction
for the fate of the main-sequence star remnant during subsequent
encounters is very difficult to obtain from our simulations and that
more polished numerical and/or method that can easily handle such a long period
of hydrodynamic simulations is neccessary for that.

\subsection{The Disk}

Unlike the binary system where mass
transfer to a disk from outside the Roche lobe of a companion is
relatively stable, the formation of an accretion disk following a tidal
disruption is so abrupt that too much material is put in the disk in a short
amount of time.  Consequently, the mass accretion rate from the disk onto the
neutron star is much higher than the Eddington limit.
Verbunt \etal (1987) estimated the amount of accreted
mass to be only ($\dot M_{Edd} / \dot M) M_d$ which is insufficient to lead to
millisecond periods, where $\dot M_{Edd}$ is the mass accretion rate
corresponding to the Eddington limit luminosity, $\dot M$ the mass accretion
rate onto the neutron star, and $M_d$ the disk mass.
 
The spin period $P_s$ of the neutron star after the accretion is determined by
total amount of accreted mass $\Delta M_{acc}$, and the the relationship between
the final $P_s$ and $\Delta M_{acc}$ can be written as
 
\begin{equation}
\Delta M_{acc} \simeq 0.20 f\left( {P_s \over {\rm ms}} \right) ^{-4/3} \msun
\label{deltam}
\end{equation}
( Phinney \& Kulkarni 1994), where f $\gsim 1$ is a factor that depends on the details of the accretion, and
the moment of inertia of the neutron star is assumed to be $10^{45} {\rm g
\, cm^2}$.   Until the neuron star reaches its equilibrium period,
$f$ remains to be around 3 (Ghosh \& Lamb 1992).
Further accretion after the equilibrium period does not spin up the star since
$f$ has become much higher than 1.

If the steady, thin disk analysis by Ghosh
\& Lamb (1992) with $f \simeq 3$ is still applicable
and if the estimation of mass transferred to the neutron star by
Verbunt \etal (1987) is valid, the accretion followed by tidal disruption
with a rate of $1000 \, \dot M_{Edd}$ may be able to accelerate $P_s$ to
$\gsim 100$ ms.

On the other hand, if there is a way for all disk material to overcome the
Eddington limit and to accrete onto the neutron star's surface, the
accretion would be able to accelerate $P_s$ to $\sim$ 1 ms.

The reality will reside somewhere between the above two extreme cases, and
the authors will  pursue a study on this heavy accretion disk around a neuron
star as a next step of the present study.

\section{DISCUSSION}
 
We have investigated the consequences of close encounters between a
neutron star and a main-sequence star using the SPH method.  We have found
that the mass fraction stripped from the main-sequence star during the
encounter, $\Delta m$, is determined only by a single parameter
$\eta^{\prime}$ in the ranges of $m$ and $v_\infty$ used in our
simulations.  The orbital energy deposited in the stellar
envelope per unit mass, $\Delta E/m$ seems responsible for the mass
stripping because it also can be well approximated to be a function of
$\eta^{\prime}$.
 
From simulations with relatively large $\eta^{\prime}$,
the core contraction of the main-sequence star remnant
has been witnessed as well as the envelope expansion.  The radiation
from the contracted core will help the envelope expansion and the envelop
may not be able to restore itself to the original place.
 
The mass of the disk formed around a neutron star after an encounter
is as much as of order 0.1 $\msun$.  If all of this mass could
be accreted onto the neutron star's surface, the accretion could
accelerate the spin of the neutorn star down to 1 ms.  However,
this kind of heavy accretion disks are subject to high radiation
pressure and considerable fraction of the disk is believed to
be blown away, which is against the spin-up of the neutron star.
The fate of these heavy accretion disks are uncertain
and remains to be studied intensively, to explain the high abundance
of isolated, recycled pulsars in globular clusters.

\acknowledgements
We are grateful to Dr. M-G Park for useful conversation on the
accretion disk theory.
This research was supported in part by the Korea Research Foundation through
non-directed research funds in 1993.

\vfill\eject

\begin{planotable}{rrrrrrrrrr}
\tablewidth{500pt}
\tablecaption{Simulation Parameters}
\tablehead{
\colhead{Simulation} & \colhead{$M_{MS}$} & \colhead{$r_{\min}$} &
\colhead{$\eta$} & \colhead{$\eta^{\prime}$} &
\colhead{Simulation} & \colhead{$M_{MS}$} & \colhead{$r_{\min}$} &
\colhead{$\eta$} & \colhead{$\eta^{\prime}$} \\[.2ex]
\colhead{} & \colhead{($\msun$)} & \colhead{($10^{10}$ cm)} & \colhead{} &
\colhead{} &
\colhead{} & \colhead{($\msun$)} & \colhead{($10^{10}$ cm)} & \colhead{} &
\colhead{}}
\startdata
m2p02 & 2 & 2 & 0.60 & 0.44 & m5p07 & 5 &  7 & 1.45 & 1.22\nl
m2p03 & 2 & 3 & 1.11 & 0.80 & m5p09 & 5 &  9 & 2.16 & 1.78\nl
m2p04 & 2 & 4 & 1.71 & 1.23 & m5p11 & 5 & 11 & 2.86 & 2.41\nl
m2p05 & 2 & 5 & 2.39 & 1.72 & m7p05 & 7 &  5 & 0.60 & 0.53\nl
m2p06 & 2 & 6 & 3.14 & 2.27 & m7p08 & 7 &  8 & 1.20 & 1.07\nl
m3p04 & 3 & 4 & 1.10 & 0.85 & m7p11 & 7 & 11 & 1.94 & 1.73\nl
m5p06 & 5 & 6 & 1.15 & 0.97 & m7p15 & 7 & 15 & 3.09 & 2.75\nl
\end{planotable}
 
\begin{planotable}{rrrr}
\tablewidth{300pt}
\tablecaption{Orbital Elements of Selected Simulations}
\tablehead{
\colhead{Simulation} & \colhead{$r_{\min}$} & \colhead{$r_{\min,2}$
\tablenotemark{a}} & \colhead{$P$} \\[.2ex]
\colhead{} & \colhead{($10^{10}$ cm)} & \colhead{($10^{10}$ cm)} &
\colhead{(h)}}
\startdata
m7p08 &  8 & 10.9 &    23\nl
m7p11 & 11 & 18.1 &   210\nl
m7p15 & 15 & 22.4 & 28000\nl
\tablenotetext{a}{Perigee distance at the second encounter}
\end{planotable}

\vfill\eject

\vfill\eject

\begin{figure}
\plotone{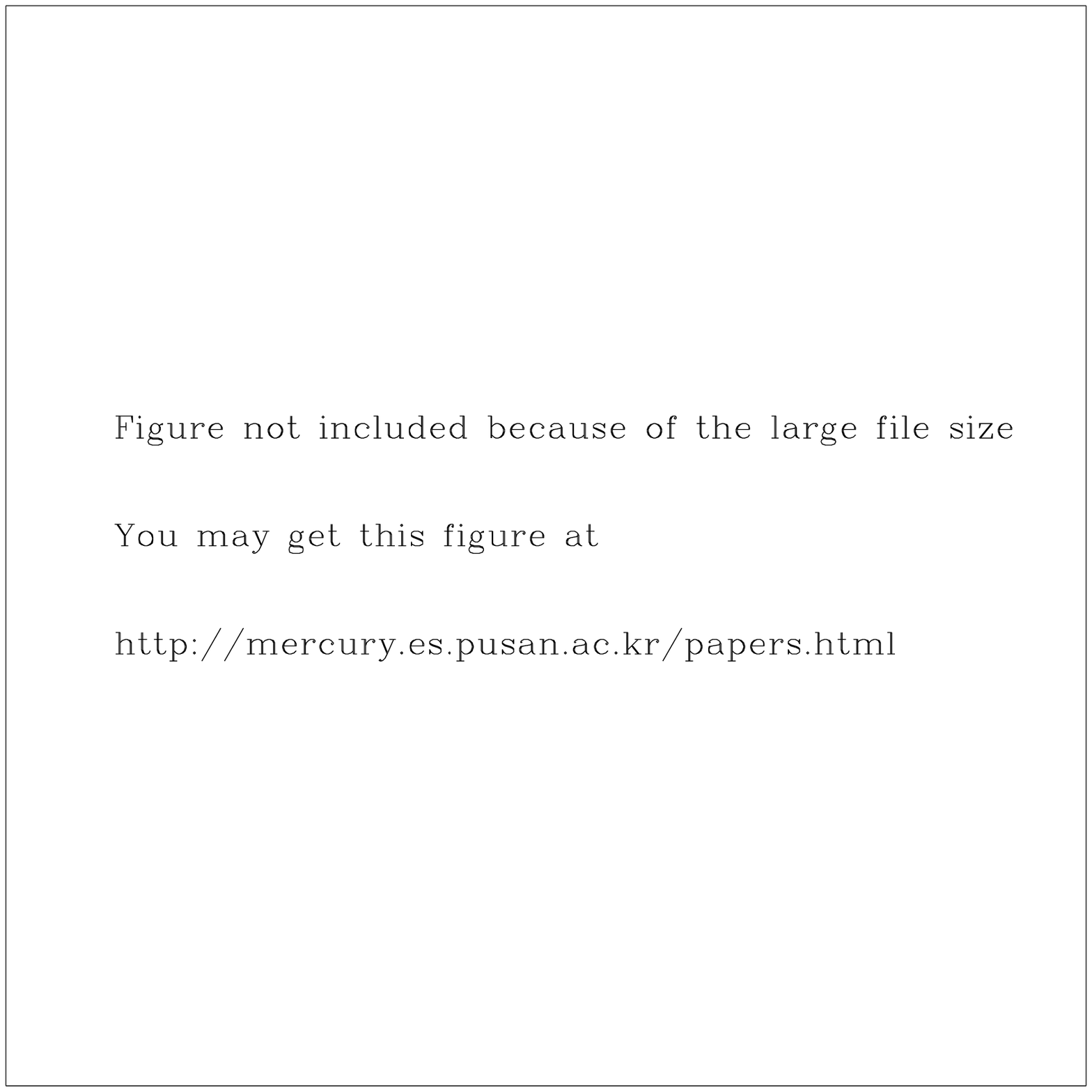}
\caption{
Four snapshots of simulation m7p08 with $M_{MS}=0.7 \msun$ and
$r_{\min}=7 \times 10^{10}$ cm.  The hyperbola is the trajectory that the
main-sequence star would follow if it were a point mass.  All snap shots are
in neutron-star-relative coordinates such that the neutron star represented
by a cross resides always at (0,0).  The axes are in units of $R_{MS}$.
}
\end{figure}

\begin{figure}
\plotone{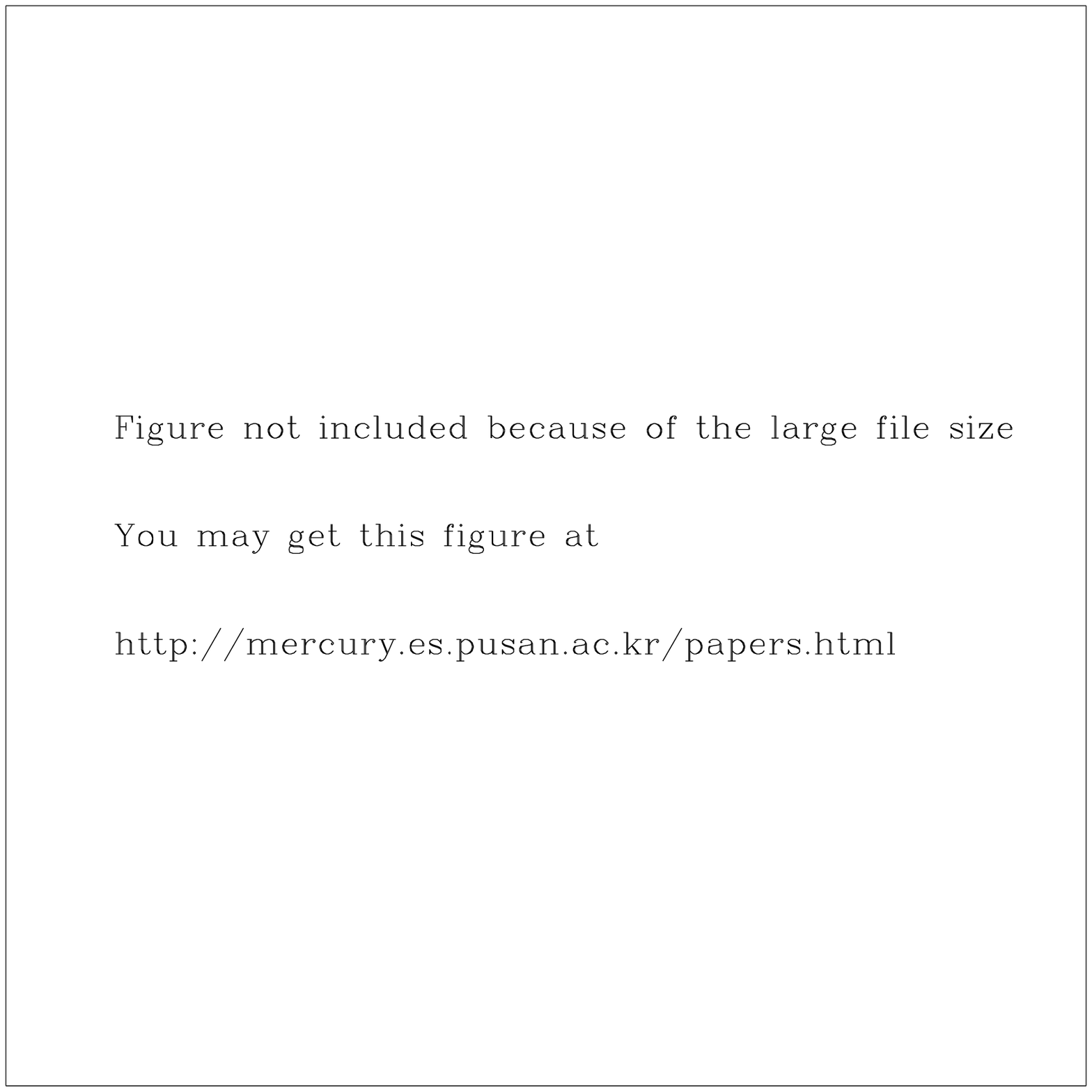}
\caption{
Reproduction of the last snapshot in Figure 1.  Small dots are the
particles bound to the neutron star, large dots those bound to the
main-sequence star, and $\times$ those unbound to the system.  For membership
criteria, see text.  The axes are in units of $R_{MS}$.
}
\end{figure}

\begin{figure}
\plotone{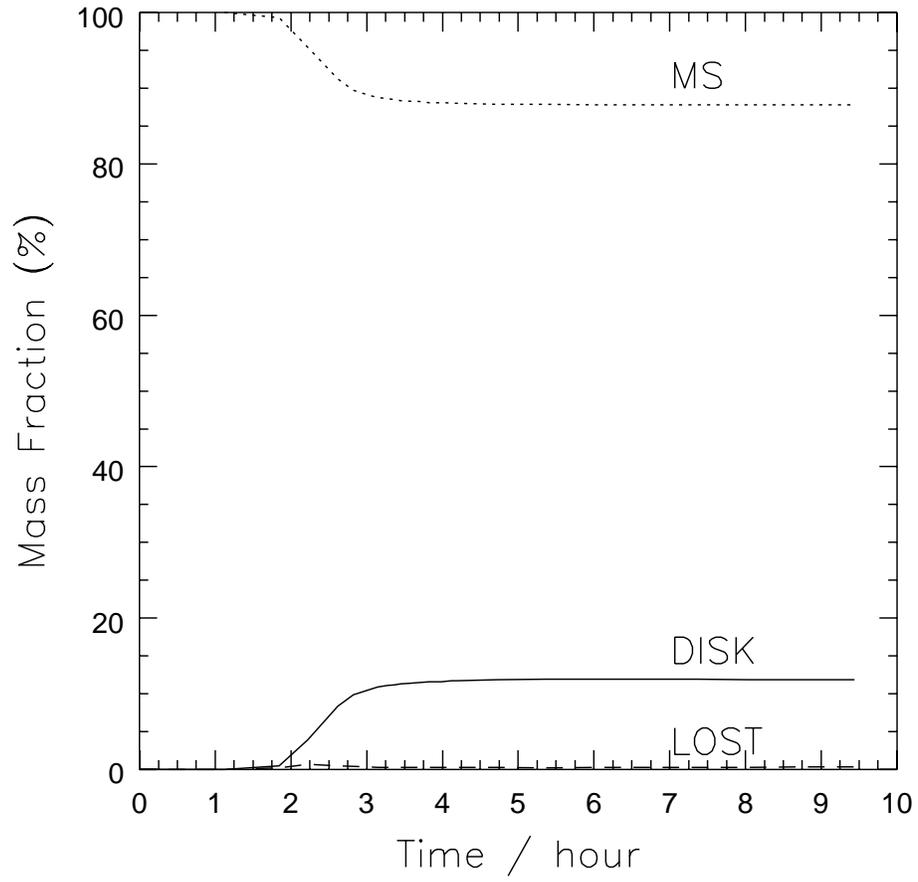}
\caption{
Temporal evolution of mass fractions that are bound to the
neutron star (DISK), bound to the main-sequence star (MS), and unbound to the
system (LOST), for simulation m7p08.
}
\end{figure}

\begin{figure}
\plotone{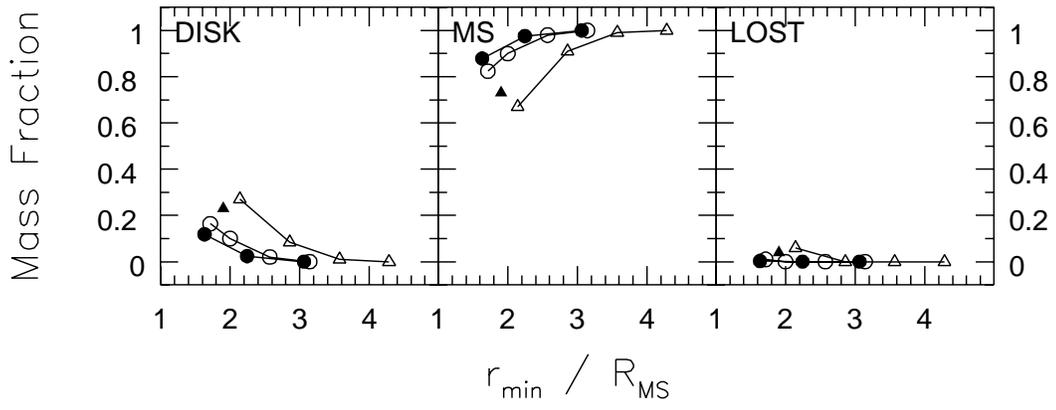}
\caption{
Mass fractions that are bound to the neutron star (DISK), bound
to the main-sequence star (MS), and unbound to the system (LOST) from the
last dumps of all 14 simulations.  Filled circles are for $0.7\msun$, open
circles for $0.5\msun$, filled triangle for $0.3\msun$, and open triangle
for $0.2\msun$ main-sequence stars.
}
\end{figure}

\begin{figure}
\plotone{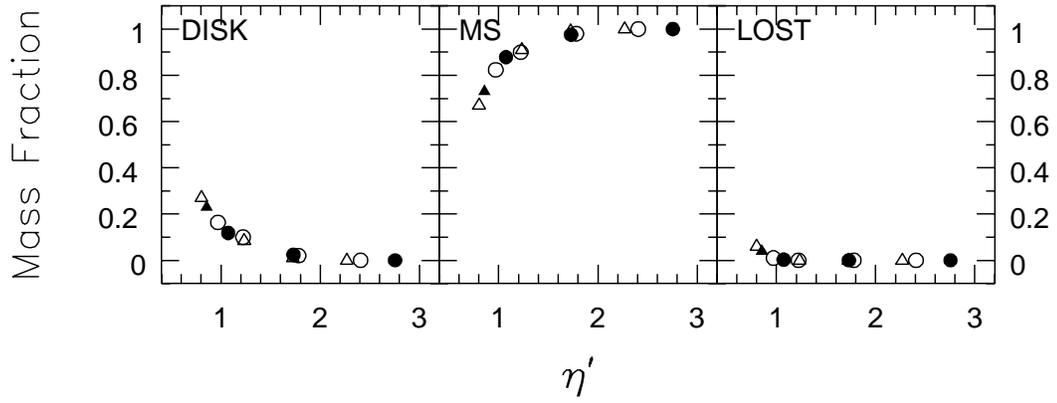}
\caption{
Mass factions.  The abscissa of Figure 4 is rescaled for
$\eta^{\prime}$ now.  Symbols are the same as in Figure 4.
}
\end{figure}

\begin{figure}
\plotone{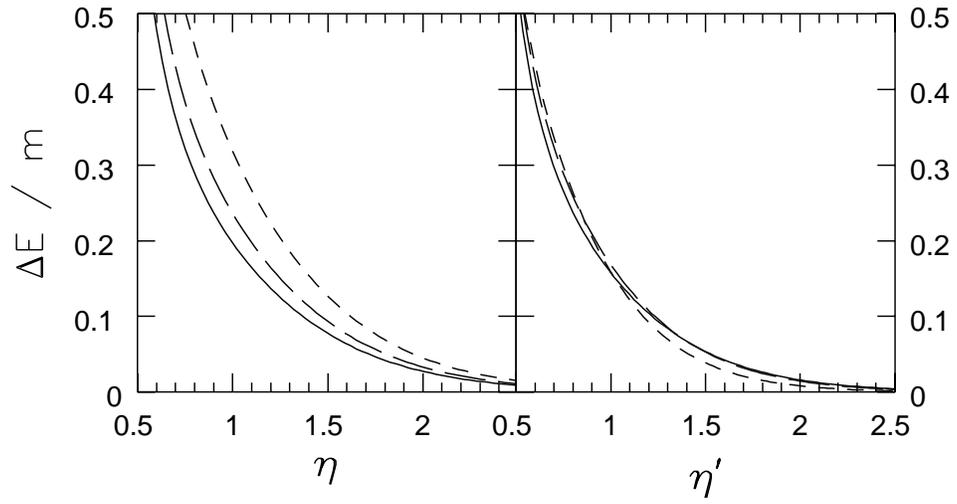}
\caption{
Theoretical expectation of orbital energy deposited in the stellar
envelope during the first encounter per stellar mass as functions of $\eta$ and
$\eta^{\prime}$ for 0.7$\msun$ (solid), 0.5$\msun$ (long-dashed), and
0.2$\msun$ (short-dashed).  $\Delta E/m$ in units of ${\rm
GM_{\odot}/R_{\odot}}$.
}
\end{figure}

\begin{figure}
\plotone{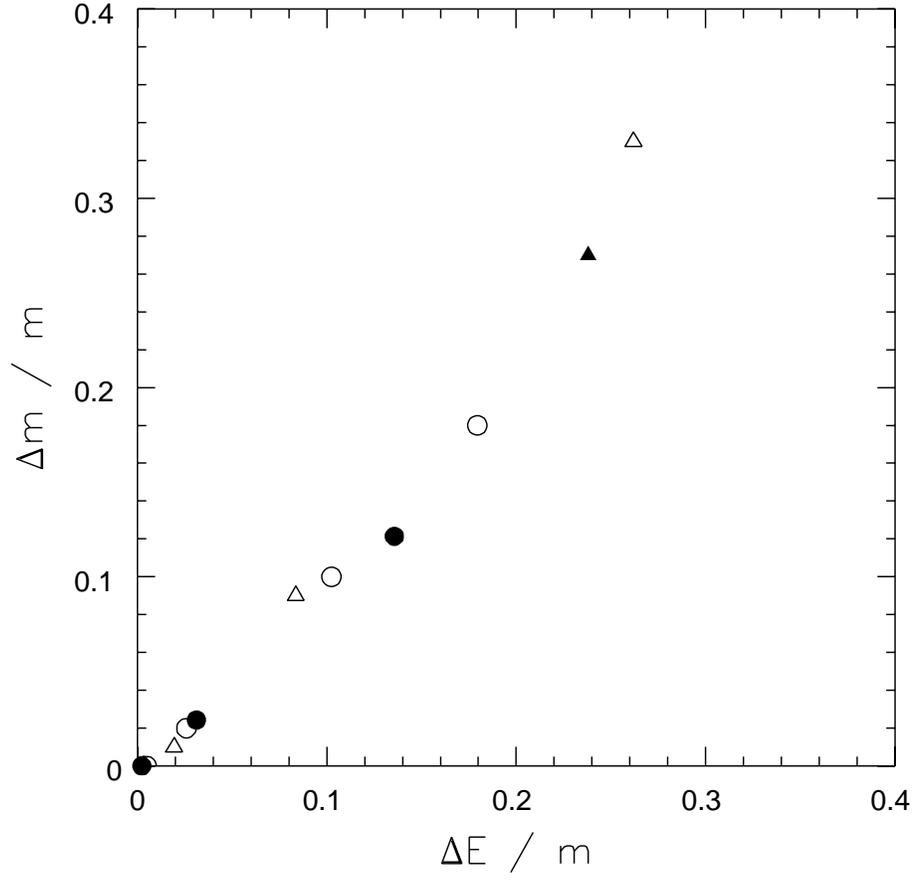}
\caption{
The amount of mass stripped per stellar mass versus that
of orbital energy deposited in the stellar envelope per stellar mass during
the first encounter.  $\Delta E/m$ in units of ${\rm GM_{\odot}/R_{\odot}}$.
Symbols are the same as in Figure 4.
}
\end{figure}

\begin{figure}
\plotone{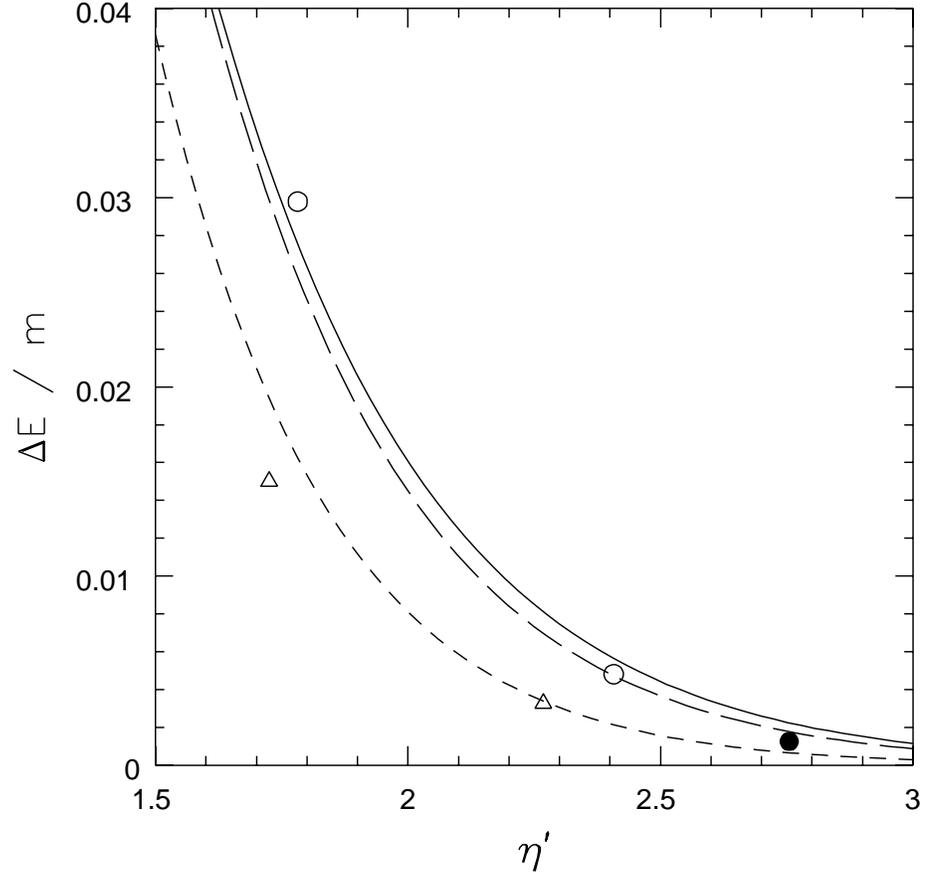}
\caption{
Comparison of the amount of orbital energy deposited in stellar
envelope per stellar mass between the theory and our simulations.
$\Delta E/m$ in units of ${\rm GM_{\odot}/R_{\odot}}$.  Solid line: $0.7\msun$;
long-dashed line: $0.5\msun$; short-dashed line: $0.2\msun$.
Symbols are the same as in Figure 4.
}
\end{figure}

\begin{figure}
\plotone{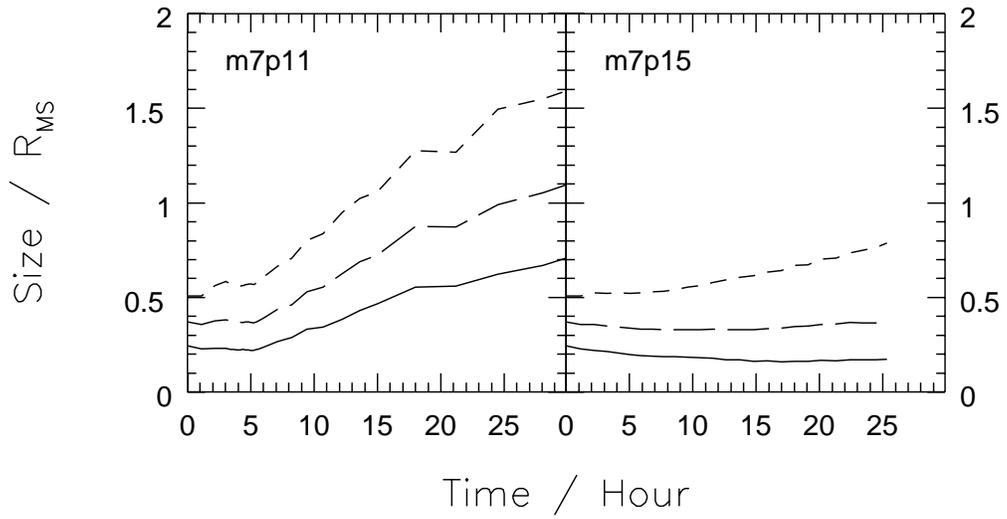}
\caption{
Expansion of the main-sequence star remnants of simulations
m7p11 and m7p15.  Solid lines are the sizes that encompass 25 \% of the mass
that satisfy the criterion B.  Long-dashed lines for 50 \% and short-dashed
lins for 75 \%.  The ordinate in units of the original size of the
main-sequence star, $0.7 \msun$.
}
\end{figure}

\end{document}